\documentclass[aps,12pt]{revtex4}
\usepackage{graphicx}
\newcommand{\be}{\begin{equation}}
\newcommand{\ee}{\end{equation}}
\newcommand{\bea}{\begin{eqnarray}}
\newcommand{\eea}{\end{eqnarray}}
\newcommand{\beas}{\begin{eqnarray*}}
\newcommand{\eeas}{\end{eqnarray*}}
\newcommand{\bd}{\begin{displaymath}}
\newcommand{\ed}{\end{displaymath}}
\def\shiftleft#1{#1\llap{#1\hskip 0.04em}}
\def\shiftdown#1{#1\llap{\lower.04ex\hbox{#1}}}
\def\thick#1{\shiftdown{\shiftleft{#1}}}
\def\b#1{\thick{\hbox{$#1$}}}

\begin{document}
\baselineskip 0.555cm

\centerline{\Large {\bf Axial exchange currents and nucleon spin} } 

\vskip1cm

\centerline{\large D. Barquilla-Cano$^1$,  A. J. Buchmann$^2$ 
and E. Hern\'andez$^1$}
\vskip 1.2 cm 
\begin{center}
$^1$ Grupo de Fisica Nuclear, Facultad de Ciencias, Universidad 
de Salamanca  \\
Plaza de la Merced s/n, E-37008 Salamanca, Spain\\
$^2$ Institut f\"ur Theoretische Physik, Universit\"at T\"ubingen\\
Auf der Morgenstelle 14, D-72076 T\"ubingen, Germany \\
\end{center}
\vskip 1.2 cm
\centerline{\Large Abstract}

\noindent
We calculate the axial couplings $g_A^8(0)$ and $g_A^0(0)$
related to the spin of the nucleon
in a constituent quark model. 
In addition to the standard one-body axial currents, 
the model includes two-body axial exchange currents.
The latter are necessary to satisfy the Partial Conservation of Axial 
Current (PCAC) condition.
For both axial couplings we find significant corrections to the standard 
quark model prediction.  Exchange currents reduce the valence quark 
contribution to the nucleon spin and afford an 
interpretation of the missing nucleon spin as orbital angular momentum
carried by nonvalence quark degrees of freedom. 


\noindent
\section{Introduction}
\label{section:introduction}
\nobreak

The question which degrees of freedom carry which part 
of the nucleon's spin has stimulated a great deal of research 
on both the experimental and theoretical side~\cite{seh74}.
In a quantum field theoretical description~\cite{Ji} 
the nucleon spin $J$ is built from the quark 
spins $\Sigma$, quark orbital angular 
momentum $L_q$, gluon spin $\Delta g$, and from gluon orbital angular 
momentum $L_g$ as
\be
\label{spindecomp}
J= \frac{1}{2} \Sigma + L_q + \Delta g + L_g =\frac{1}{2} .
\ee
Presently, there is still considerable controversy over how 
the nucleon spin $J=1/2$ is made up in detail.
In the standard constituent quark model~\cite{clo92}, 
which uses only one-body 
axial currents, one obtains a clear answer, namely $J=\Sigma/2=1/2$, i.e.,
the nucleon spin is the sum of the three constituent quark spins and nothing 
else. On the other hand, in the Skyrme model~\cite{bro88} 
most of the nucleon spin is due to orbital angular momentum of the quarks.
Experimental results~\cite{abe98,emc} for the quark spin sum  
indicate that $\Sigma/2=0.16(5)$, i.e., 
only $1/3$ of the total angular momentum of the nucleon is due to quark spins.
This marked disagreement between the standard constituent quark model 
and experiment has often been interpreted as a severe shortcoming of this 
model (``spin crisis''). 

However, the simple additive constituent quark model violates the partial 
conservation of the axial current (PCAC) condition, which is an important 
symmetry constraint that must be satisfied in a consistent theory. The purpose 
of this paper is to show that if the constituent quark model
is supplemented by two-body axial exchange currents so that the PCAC condition 
is satisfied, the valence quark contribution to the nucleon spin 
is considerably reduced. Furthermore, the model suggests that the nucleon
contains a substantial amount of orbital angular momentum carried by
nonvalence quark degrees of freedom effectively described by exchange
currents.

The paper  is organized as follows.
In sect.~\ref{section:dis} we review the relation between the axial 
current matrix elements and the spin fraction carried by quarks.
The axial current operators of the
constituent quark model are briefly discussed in 
sect.~\ref{section:quarkmodel}. 
Our results for the nucleon spin structure are
presented in sect.~\ref{section:results} and summarized  
in sect.~\ref{section:summary}.

\section{Spin structure of the nucleon}

\label{section:dis}

Experimentally, information on the spin structure of the nucleon is 
obtained from deep inelastic scattering (DIS) of polarized leptons 
on longitudinally 
polarized protons. In such an experiment one is measuring the 
spin-dependent proton structure function $g_1^p(x,Q^2)$, which depends 
on the four-momentum  transfer $Q^2=-q_{\mu}\, q^{\mu}$ 
and the Bjorken scaling variable  $x=Q^2/(2 M\nu)$~\cite{note0}.
At $Q^2$ and $\nu$ large compared to the nucleon mass $M$ (scaling limit), 
the integral of the spin structure function can be calculated 
in the framework of QCD to first order in the strong coupling constant 
$\alpha_s$~\cite{carlitz} 

\be
\label{firstmoment}
\label{Gamma1}
\Gamma_1^p(Q^2)=: \int_0^1 g_1^p(x,Q^2)\ dx =
\left (\frac{4}{18} \Delta u(Q^2)
 + \frac{1}{18} \Delta d(Q^2)
+\frac{1}{18} \Delta
s(Q^2) \right ) \bigg(1-\frac{\alpha_s(Q^2)}{\pi}\bigg),
\ee
where the 
$\Delta q(Q^2)$ are the fractions of the nucleon spin carried 
by the quarks and antiquarks of flavor $q=(u, d, s)$ evaluated at the
renormalization scale $\mu^2=Q^2$.
In the parton model, the spin fractions 
carried by the individual quark and antiquark flavors are defined as  
\begin{eqnarray}
\label{partonmodel}
\Delta q(Q^2)=\int_0^1 dx\, 
\big(q \! \uparrow (Q^2,x)+\bar{q}\!\uparrow  (Q^2,x)
-q \!\downarrow  (Q^2,x)-\bar{q} \!\downarrow  (Q^2,x) \big).
\end{eqnarray}
The quark momentum distributions 
$q\! \uparrow \!(Q^2,x)$ and $q\!\downarrow \!(Q^2,x)$ denote
the probability for finding in the nucleon a current or QCD quark~\cite{note1} 
of flavor $q$ with momentum fraction $x$ of the 
total nucleon momentum and spin parallel or 
antiparallel to the proton spin.
The spin quantization axis is chosen along the proton momentum.
The quark momentum distributions, e.g., $q \!\uparrow   (Q^2,x)$, 
can be further decomposed:
\be
\label{valseadecomp}
q \!\uparrow  (Q^2,x)=q_{val} \!\uparrow   (Q^2,x)+q_{sea} \!\uparrow  (Q^2,x)
\ee 
where $q_{val} \!\uparrow  (Q^2,x)$  and $q_{sea} \!\uparrow  (Q^2,x)$
is the contribution of the QCD valence and sea quarks. 
Note that the definition of $\Delta q$ in Eq.(\ref{partonmodel}) 
also contains the antiquark distributions denoted by 
$\bar{q} \!\uparrow (Q^2,x)$ and $\bar{q} \!\downarrow (Q^2,x)$.
Hence, the spin fractions involve both valence quarks and 
sea quark-antiquark pairs.

The fraction of the nucleon spin fractions 
carried by the individual quark flavors $q$ 
can be expressed as axial vector current matrix elements
\be
\label{acme}
\Delta q(\mu^2) =
\langle p \!\uparrow \left \vert \,  g_{Aq} \, \left ({\bar q} \, 
\gamma_{3} \, \gamma_5 \, q \right )\vert_{\mu^2} 
\right \vert p \!\uparrow \rangle, 
\ee
where $\gamma_3=\gamma_z$ and $\gamma_5$ are Dirac matrices, $\mu^2$ 
refers 
to the mass scale at which the axial current operator is renormalized,
and $g_{Aq}$ is the quark axial current coupling constant. 

In the constituent quark model one does a tree level calculation with the
usual assumption 
that the quark fields $q$ in Eq.(\ref{acme}) are described by free 
Dirac spinors. In addition, the static limit is taken so that 
the contribution from the lower components of the Dirac spinors can be 
neglected. Furthermore, it is assumed that $g_{Aq}=1$, i.e., the same as 
for structureless QCD quarks, in which case 
the axial current matrix element reduces to
\be
\label{nonrelativistic}
\Delta q
= \langle p \!\uparrow \vert  \sigma_z^q \, \vert \, p \!\uparrow  \rangle.
\ee
Thus, in this approximation,
the axial quark current is the one-body~\cite{note2}
quark spin operator $\sigma_{z}^q$.
Its matrix element between nucleon states measures the contribution of 
a valence quark with flavor $q$ to the nucleon spin, where $q$ 
is restricted to the $u$ and $d$ flavors.
The total angular momentum of the nucleon is then simply the sum of 
the $u$ and $d$ quark spin fractions.
With the usual SU(6) spin-flavor wave function of the proton 
and the one-body axial current operator
in Eq.(\ref{nonrelativistic}),
one obtains for the spin fractions
$ \Delta u= 4/3$ and $\Delta d= -1/3$ adding up to
\be
\label{additivequarkmodel}
<J_z>=<S_z>=\frac{1}{2} \, (\Delta u+\Delta d)=\frac{1}{2}.
\ee

More generally, allowing for three flavors in the axial current operator 
in Eq.(\ref{acme}), and evaluating matrix elements for SU(3) flavor octet 
baryons, one finds that the three axial 
form factors at zero momentum transfer, 
$g_A(0)$, $g_A^8(0)$, and $g_A^0(0)$ 
are the relevant 
quantities that carry the 
information on the nucleon's spin structure. The relations between
the axial couplings $g_A$ and the spin fractions $\Delta q$ are
\goodbreak
\bea 
\label{axialffspinfrac}
g_A(0) & = & \Delta u(Q^2) - \Delta d(Q^2)  \nonumber  \\
g_A^8(0) & = & \Delta u(Q^2) + \Delta d(Q^2)- 2 \Delta s(Q^2)  
\nonumber  \\
g_A^0(0)_{Q^2} & = & \Delta u(Q^2) 
+ \Delta d(Q^2) +  \Delta s(Q^2) =:\Sigma. 
\eea
The flavor octet $g_A(0)$ (isovector) and $g_A^8(0)$ 
(hypercharge) couplings are independent of the renormalization 
point. On the other hand, the flavor singlet axial coupling $g_A^0(0)$ depends 
on the renormalization scale at which it is measured~\cite{comment0},
hence the subscript $Q^2$. 

Experimental results for $g_A(0)$ and $g_A^8(0)$ have been obtained from 
the weak semileptonic decays 
of octet baryons~\cite{comment1,pdg,lipkin,flores}.
With $\Gamma_1^p(Q^2)$ measured in DIS~\cite{abe98} 
and the axial couplings $g_A(0)$ and $g_A^8(0)$ determined 
from $\beta$ decays, there are three experimental data and 
three unknown spin fractions $\Delta q$. 
The combined DIS and hyperon $\beta$-decay data 
give~\cite{abe98} 
\begin{eqnarray}
\label{deltaq}
\Delta u = \hspace{.35cm} 0.83 \pm 0.03,\hspace{.5cm} 
\Delta d =-0.43 \pm 0.03, \hspace{.5cm}
\Delta s = -0.09 \pm 0.04.
\end{eqnarray} 
The sum of these experimental spin fractions  
$\Sigma/2=0.16(5)$ is considerably smaller 
than the additive quark model  result $\Sigma/2=1/2$ of 
Eq.(\ref{additivequarkmodel}).
However, the 
measured spin sum does not only include the 
contribution of the valence quark spins
but also the contribution of quark-antiquark pairs in the Dirac sea
as is evident from the definition of $\Delta q$ in  Eq.(\ref{partonmodel}).
These degrees of freedom are not properly included in the 
additive quark model in which the axial quark current is approximated as 
a one-body operator.

\section{The PCAC relation and axial exchange currents }

\label{section:quarkmodel}

When discussing the failure of the additive constituent quark model 
in correctly describing the axial couplings of the nucleon 
and the related spin fractions carried by the valence and nonvalence 
degrees of freedom, 
it is usually not mentioned that this model, which 
uses only one-body axial currents, violates the PCAC relation~\cite{david} 
\be
\label{pcac}
{\bf q} \cdot {\bf A}({\bf q}) -   
[H,A^0({\bf q})]
= -i\ \sqrt2\ f_{\pi}\frac{m_{\pi}^2}{q^2-m_{\pi}^2} M^{\pi}({\bf q}).
\ee
This can be seen as follows.
The PCAC relation links the strong interaction Hamiltonian
$H$, the weak axial current $A^{\mu}=(A^0,{\bf A})$ operators, 
and the pion emission operator described by $M^{\pi}$. Here,
$m_{\pi}$ is the pion mass and $f_{\pi}$ 
is the pion decay constant for which we use $f_{\pi}=93$ MeV.
If the quark Hamiltonian contains two-body potentials, 
(e.g. due to gluon exchange), which do not commute with the axial 
charge density $A_0$, the PCAC relation
demands that there be corresponding two-body axial currents ${\bf A}$
(e.g. axial gluon exchange currents) 
and two-body emission operators to counterbalance the 
contribution from the two-body potentials in the Hamiltonian~\cite{david}. 
This is analogous to the requirement of 
the continuity equation for the electromagnetic 
quark current~\cite{BHY94}, which enforces the presence of
two-body exchange currents if the two-body potentials do not commute with 
the quark charge operator.
\begin{figure}
{\resizebox{10 cm}{6.0 cm}{\includegraphics{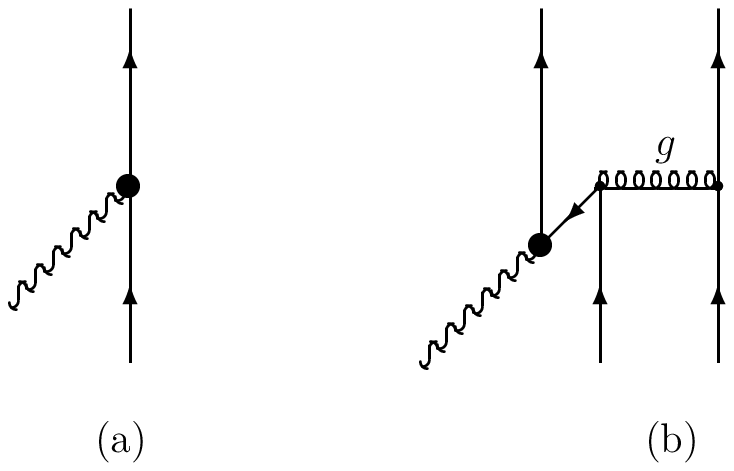}}}
\vspace{-1.2 cm}
\caption{\label{fig:Fig1} Feynman diagrams for the axial current 
operators of constituent quarks:
(a) one-body axial current, (b) two-body gluon exchange axial current.
The wavy lines correspond to weak gauge bosons coupling to the axial 
currents. The black dot denotes the axial quark coupling.}
\end{figure}

Eq.(\ref{pcac}) leads to the  Goldberger-Treiman 
relation~\cite{Eri88,comment3} between the quark axial coupling $g_{Aq}$, 
and the pion-quark coupling constant $g_{\pi q}$
\be
\label{gaq}
g_{Aq}=f_{\pi}\, \frac{g_{\pi q}}{m_q}.
\ee
Here, $m_q$ is the constituent quark mass.
In a recent work~\cite{david} we have shown that a good description of  
the nucleon's isovector axial vector form factor $g_A(q^2)$
is obtained if the axial current operator ${\bf A}$ satisfies the 
PCAC condition~(\ref{pcac}), in particular
if $g_{Aq}$ is determined from Eq.(\ref{gaq}).

In the present paper, we generalize these results to three flavors, 
and calculate the axial form factors $g^8_A(0)$ and 
$g^0_A(0)$ for SU(3) flavor octet baryons in a chiral constituent quark model
based on a non-linear $\sigma$-model 
Lagrangian~\cite{manohar} that is invariant under
U(3)$_V \times $U(3)$_A$ chiral transfomations.
After an expansion in
powers of $1/f_{\pi}$ the effective Lagrangian reads 
\bea
\label{lag}
{\cal L}= & &\overline{\Psi}(i\gamma^{\mu}\partial_{\mu}-M')
\Psi + \frac{1}{2}\,\partial^{\mu}\Phi_j\partial_{\mu}\Phi_j
+ \frac{1}{2}\,\partial^{\mu}\Phi_0\partial_{\mu}\Phi_0 \nonumber \\
&&\hspace{-0.5cm}-\frac{1}{2}\,m^2_{\Phi_j}\Phi^2_j 
 -\frac{1}{2}\,m^2_{\Phi_0}\Phi^2_0 
-\frac{1}{2}\,m^2_{\Phi_{8,0}}\Phi_8 \Phi_0 
-\frac{3\,a}{2\, N_c}\,\Phi_0^2\nonumber \\ 
&&\hspace{-0.5cm} 
-\frac{1}{2}\ tr \left( F_{\mu\nu}\, F^{\mu\nu}\right)
-\,g\overline{\Psi}\gamma^{\mu}\,G_{\mu}\Psi 
+\frac{g_{Aq}}{2f_{\pi}}\,\overline{\Psi}\gamma^{\mu}\gamma_5\,\b{\lambda}^j
\Psi\ \partial_{\mu}\,\Phi_j
+\frac{g^0_{Aq}}{2f_{\pi}}\,\overline{\Psi}\gamma^{\mu}\gamma_5\,{\lambda}_0
\Psi\ \partial_{\mu}\,\Phi_0 \nonumber \\
&&\hspace{-0.5cm} 
-\frac{1}{4f^2_{\pi}}\, f_{jkl}\,\overline{\Psi}
\gamma^{\mu}\,\b{\lambda}^j\Psi\,\Phi_k \partial_{\mu}\Phi_l.
\eea
Here, $\Psi$ and $\Phi_0$, $\Phi_j$ with $j=1 \cdots 8$ 
represent the quark and the nonet of pseudoscalar meson fields~\cite{comment11}
respectively, $G_{\mu}$ is the gluon field, and $F_{\mu\nu}$ is the gluon field
strength tensor. The $\b{\lambda}^j$ are SU(3) flavor matrices, 
$\lambda^0=\sqrt{2/3}\,\, {\bf 1} $, 
and the $f_{jkl}$ are the SU(3) structure constants. 

Due to the complicated structure of the QCD vacuum 
characterized by a nonzero quark condensate 
$<0\mid \Psi{\bar \Psi} \mid 0> \ne 0$,
the chiral symmetry of the
Lagrangian is spontaneously broken. This gives rise to nonzero 
constituent quark mass terms. In addition, there are small QCD quark mass terms
which break chiral symmetry explicitly.
The matrix $M'=diag(m_q, m_q, m_s)$ in Eq.(\ref{lag}) 
contains both contributions. 
The small QCD quark mass terms together with a nonzero quark condensate 
are also responsible for the 
meson mass terms $m_{\Phi_0}^2,\ m_{\Phi_j}^2\ (j=1\cdots 8)$, and
$m_{\Phi_{8,0}}^2$.

The axial or chiral U(1)$_A$ anomaly of QCD is also taken into account 
in this effective low energy
Lagrangian through an extra mass term $-3a/2N_c $ for the $\Phi_0$,
where $N_c$ is the number of colors and $a$ is a ${\cal O}(N_c^0)$ 
quantity having dimensions of mass square. The axial anomaly 
breaks U(3)$_F$ flavor symmetry and thus gives rise to different flavor octet
and singlet meson masses and axial coupling constants.
To take SU(3) flavor symmetry breaking into account, 
we allow that the hypercharge axial quark coupling $g^8_{Aq}$ 
is different from the isovector axial quark coupling $g_{Aq}$ (see below).

From the fundamental interaction vertices of the non-linear 
$\sigma$ model Lagrangian we have derived 
the potentials and axial current operators of the chiral
constituent quark model.
The resulting quark potential model contains a gluon exchange potential
with strength $\alpha_s=g^2/4\pi$,  
Goldstone boson exchange interactions with coupling constants
$g_{Aq}/2f_{\pi}$, a confinement interaction, and the corresponding 
axial currents, as discussed in detail in Ref.~\cite{nuclth}. 
For the present 
purpose it is sufficient to list only the one-body and gluon exchange 
axial current operators,
which contribute to the flavor octet and flavor singlet axial couplings
$g_A^8(0)$ and $g_A^0(0)$. These operators are derived from the Feynman 
diagrams in Fig.~\ref{fig:Fig1}.
The one-body isovector axial current operator reads
\bea
\label{one-body isovector}
{\bf A}^j_{imp}({\bf q}) &=& g_{Aq}\ \sum_k\ e^{i{\bf q}\cdot{\bf r}_k} 
\ \frac{\b{\tau}_k^{j}}{\sqrt2}
\ \b{\sigma}_k ,
\eea
where $k=1,2,3$ sums over the three constituent quarks in the nucleon,
and $j=1,2,3$ denotes the (cartesian) components of the isospin   
operator $\b{\tau}_k$. The quark spin and position operators are denoted 
by $\b{\sigma}_k$ and ${\bf r}_k$, 
and ${\bf q}$ is the three-momentum transfer 
imparted by the weak gauge boson. 
  
To generalize the SU(2) isospin axial current to 
the SU(3) flavor axial current,
we replace the isospin operator $\b{\tau}_k^{j}$ by the Gell-Mann flavor 
operator $\b{\lambda}_k^{j}$, where $j=1, \cdots, 8$ is the flavor index. 
In particular, the $j=8$ flavor (hypercharge) axial current is obtained 
by replacing $\b{\tau}_k^{j}/\sqrt{2}$  in Eq.(\ref{one-body isovector}) by 
$\sqrt{3}\,\b{\lambda}_k^8$ and $g_{Aq}$ by $g^8_{Aq}$
\bea
\label{a8i}
{\bf A}^8_{imp} ({\bf q}) &=& g^8_{Aq}\ \sum_k\ 
\ e^{i{\bf q}\cdot{\bf r}_k}\ (\sqrt{3} \b{\lambda}_k^8) \ \b{\sigma}_k.
\eea
In the constituent quark model the nucleon consists 
of $u$ and $d$ valence quarks so that we only need the left 
upper corner of the SU(3) flavor matrices 
$\b{\lambda}_k^8=1/\sqrt{3}\, diag(1, \, 1, \, -2)$, which is 
proportional to the unit matrix. Similarly, the 
flavor singlet one-body axial current is obtained by 
the replacement $g^8_{Aq} \to g_{Aq}^0$ in  Eq.(\ref{a8i})
\bea
\label{a0i}
{\bf A}^0_{imp} ({\bf q}) &\equiv& g_{Aq}^0\ \sum_k\ 
\ e^{i{\bf q}\cdot{\bf r}_k}\ \b{\sigma}_k.
\eea

From the diagram of Fig.~\ref{fig:Fig1}(b) 
we derive the gluon exchange isovector axial current~\cite{david,nuclth} 
\bea
\label{gluonexchangecurrent}
{\bf A}^j_{g}({\bf q}) &=& g_{Aq}\, \frac{\alpha_s}{16m_q^3}\, 
\sum_{k<l}
\b{\lambda}_k^c\cdot \b{\lambda}_l^c\, 
\Biggl\{ \ e^{i{\bf q}\cdot{\bf r}_k}\frac{\b{\tau}_k^{j}}{\sqrt2}\ 
\Biggl[ 
\ -i(\b{\sigma}_k\cdot {\bf r})\ {\bf q} \ \frac{1}{r^3} \nonumber \\
&&+\biggl( 3(\b{\sigma}_k+\b{\sigma}_l)\cdot 
\hat{{\bf r}}\ \hat{{\bf r}}
-(\b{\sigma}_k+\b{\sigma}_l)  \biggr)\ \frac{1}{r^3}
+\frac{8\pi}{3}(\b{\sigma}_k+\b{\sigma}_l)\ 
\delta({\bf r})\Biggr]
+(k\leftrightarrow l)\Biggr\}. 
\eea
In Eq.(\ref{gluonexchangecurrent}) 
$\b{\lambda}_k^c$ denotes the color operator of quark $k$. The relative 
coordinate between 
the two quarks exchanging a gluon is
${\bf r}= {\bf r}_k - {\bf r}_l$ with $r=\vert {\bf r} \vert$, and  
${\hat {\bf r}}={\bf r}/r$ is the corresponding unit vector.
This two-body axial exchange current is consistent with the PCAC 
constraint in Eq.(\ref{pcac})(see Ref.~\cite{david}). 

As before, 
the $j=8$ flavor component of the gluon exchange current is constructed
from the isovector current by the replacements 
$\b{\tau}_k^{j}/\sqrt{2} \to \sqrt{3}\,\b{\lambda}_k^8$ 
and $g_{Aq} \to g^8_{Aq}$ in Eq.(\ref{gluonexchangecurrent}).
Finally, the flavor singlet gluon exchange current is obtained 
from there by the substitution $g^8_{Aq} \to g_{Aq}^0$.
The hypercharge and singlet axial currents satisfy generalized PCAC 
relations~\cite{nuclth}, which are more complicated than Eq.(\ref{pcac}) 
due to the chiral anomaly of QCD. 

For an interpretation of these operators we remember that
the constituent quark model is an effective description 
of baryon structure in terms of massive valence quarks, which have the same
quantum numbers as the original QCD quarks.
Non-valence quark degrees of freedom, such as quark-antiquark pairs 
and gluons enter the model in the form of two-body currents
and renormalized quark couplings. 
Thus, the one-body axial current in Fig.~\ref{fig:Fig1}(a) 
is the current of noninteracting constituent valence 
quarks (impulse approximation), 
whereas the two-body axial current of Fig.~\ref{fig:Fig1}(b) 
describes quark-antiquark and gluon degrees of freedom resulting from the 
interaction between these valence quarks. 

Concerning the axial quark couplings for the isospin, 
hypercharge, and flavor singlet currents we recall that 
the properties of massive, spatially extended constituent quarks 
are generally different 
from those of the nearly massless and pointlike QCD quarks.  
For pointlike QCD quarks and if U(3)$_F$ flavor symmetry is exact, 
one has~\cite{Yab93}
\be
\label{qcdcouplings}
g_{Aq}=g^8_{Aq}=g^0_{Aq}=1.
\ee
In contrast, for constituent quarks, which can be viewed as QCD quarks 
surrounded by a polarization cloud of quark-antiquark pairs, the axial 
couplings is expected to be different from unity. 
In fact, several calculations~\cite{Yab93,wei91,Peris,andreas} 
show that due to the cloud of quark-antiquark pairs surrounding a QCD quark,
the isovector axial quark coupling $g_{Aq}$ is renormalized 
from 1 (QCD quark) to about 0.75 (constituent quark).
This is consistent with the value $g_{Aq} \approx 0.77$ obtained 
from the empirical pion-quark coupling constant $g_{\pi q}$ via the 
Goldberger Treiman relation in Eq.(\ref{gaq}).
In addition, because flavor symmetry is broken, one expects these 
axial couplings to be different from each other. 

As in the case of the isovector axial quark coupling $g_{Aq}$ 
we calculate the isosinglet hypercharge $g_{Aq}^8$ and the
flavor singlet $g_{Aq}^0$ axial quark couplings 
from the strong $\eta q$ and $\eta' q$ couplings and decay
constants via generalized Goldberger-Treiman relations.
Because the physical $\eta$ and $\eta'$ mesons are a mixture of the 
SU(3)$_F$ octet and singlet representations, the corresponding
Goldberger-Treiman relations are more complicated than Eq.(\ref{gaq}).
The mixing scheme~\cite{Leut98,Fel98} for the determination of $g^8_{Aq}$ 
and $g^0_{Aq}$ from $g_{\eta q}$ and $g_{\eta q'}$ involves two decay 
constants $f_8$ and $f_0$ different from $f_{\pi}$ and two mixing 
angles $\Theta_8$ and $\Theta_0$
\bea
\label{gaqfeldmann}
g_{Aq}^8 & = & \sqrt{3} \, f_{8}\  \left \{ 
\frac{\ g_{\eta q}}{m_q}\cos\Theta_8 
+ \frac{g_{\eta' q}}{m_q}\sin\Theta_8 \right \}
\nonumber \\
g_{Aq}^0 & = & \sqrt{\frac{3}{2}} \, 
f_{0} \  \left \{ -\frac{g_{\eta q}}{m_q} \sin\Theta_0 +
\frac{g_{\eta' q}}{m_q} \cos\Theta_0 \right \}.
\eea
As discussed below, the strong $\eta q$ and $\eta' q$ couplings can 
be calculated from the empirical strong $\eta N$ and $\eta' N$ couplings.
With the mixing angles and decay constants taken from Ref.~\cite{Fel98},
the axial quark couplings $g^8_{Aq}$ and $g^0_{Aq}$ can be obtained from
Eq.(\ref{gaqfeldmann}). However, it should be noted that there are different 
forms for the isosinglet Goldberger-Treiman relations in the literature, 
and there is considerable discussion concerning their validity in view of 
the chiral anomaly of QCD~\cite{Nap02,Che99,Nar99}. 

\begin{table}
\caption{ \label{table:flavorga80} 
Results for the axial couplings $g_A^8(0)$ and $g_A^0(0)$.
The individual contributions
are the one-body current (imp), the gluon exchange current (g), 
the total current (total), the experimental data (exp)~\cite{abe98}. 
For the constituent quark axial couplings we use 
$g^8_{Aq}=0.96(44)$ (second row) and $g^0_{Aq}=0.46(09)$ (fourth row) 
as determined from Eq.(\ref{gaqfeldmann}). The theoretical error bars 
in the results for $g_A^8(0)$ and $g_A^0(0)$ are due to the uncertainties
in the constituent quark axial couplings. 
}
\begin{ruledtabular}
\begin{tabular}{ccccc} 
& imp & g & total & exp \\ \hline 
$g_A^8(0)/g^8_{Aq}$  & 0.99 & -0.39 & 0.60  &  \\
$g_A^8(0)$           & 0.95($\pm 0.44$) & -0.38($\mp 0.17$) &0.58($\pm0.26$)  
&  0.58(10) \\ \hline
$g_A^0(0)/g^0_{Aq}$  & 0.99 & -0.39 & 0.60  &  \\
$g_A^0(0)$           & 0.46($\pm 0.09$) & -0.18($\mp 0.04$) & 0.28($\pm 0.05$)
& 0.33(10) \\ \hline 
\end{tabular}
\end{ruledtabular}
\end{table}

{\section{Results and Discussion}

\label{section:results}

In the following we present our results for the axial couplings 
$g_A^8(0)$ and $g_A^0(0)$ of the nucleon.
As explained in sect.~\ref{section:dis}, the nucleon 
spin fractions can be expressed in terms of these axial couplings.

\subsection{The flavor octet hypercharge axial coupling  ${\bf g_A^8(0)}$}
\label{sect:sectionga8}

The flavor octet hypercharge axial coupling $g_A^8(0)$ of the nucleon 
is given by the matrix element of the sum of one- and two-body 
axial current operators ${\bf A}^8({\bf q} \to 0)$ 
in sect.~\ref{section:quarkmodel} evaluated between 
proton wave functions.
If the Dirac sea is assumed to be SU(3) flavor symmetric, i.e.,
$
\Delta u_{sea} + \Delta {\bar u}_{sea} 
= \Delta d_{sea} + \Delta {\bar d}_{sea}
= \Delta s_{sea} + \Delta {\bar s}_{sea}
$   
this coupling measures the fraction of the proton spin that is carried by 
valence QCD quarks, namely 
$$g_A^8(0) = \Delta u_{val} + \Delta d_{val} =1$$
as in Eq.(\ref{additivequarkmodel}).  
This assumption is however not consistent with the data 
$g_A^8(0)_{exp} =0.58(10)$. Thus, in
addition to the positive valence quark contribution 
$g_A^8(0)$ contains a negative contribution coming from a 
flavor asymmetrical polarized quark-antiquark sea. 

In the first row of table~\ref{table:flavorga80}
we show our results for $g^8_{A}(0)/g^8_{Aq}$. 
The one-body 
axial current gives a contribution $A_1 \approx 1$  
because the relevant operator
${\bf A}^8_{imp}(0)/g^8_{Aq}=\sum_k \b{\sigma}_k$,  
is simply twice the total spin operator, 
and the valence quarks are mainly in relative S-wave states. 
With the two-body axial current included, we obtain
\be
\label{resultga8}
\frac{g^8_{A}(0)}{g^8_{Aq}}= (A_1+2\, A_2)=0.6. 
\ee
Here, the $A_2$ term comes from the two-body gluon exchange current.
The latter is negative and reduces the valence quark (one-body current) 
result by about 40$\%$.
Quark-antiquark degrees of freedom effectively described by the 
gluon exchange current in Fig.\ref{fig:Fig1}(b) are responsible for this
reduction. 

This result can also be derived from the general spin-flavor 
structure of the one- and two-body axial currents in the flavor 
symmetry limit~\cite{Mor89}, which for the hypercharge current reads
\be 
\label{gp}
{\bf A}^8_z= 
a \, \sum_{i=1}^3 (\sqrt{3} \, \b{\lambda}^8_i) \,  \b{\sigma}_{i\, z} 
+ b \, \sum_{i \ne j}^3 (\sqrt{3} \,  \b{\lambda}^8_i) \,  \b{\sigma}_{j\, z}.
\ee 
Here, the constants $a$ and $b$ are parameters that parametrize the 
orbital and color space matrix elements. 
For $\mid {\bf q} \mid \to 0$ 
and S-waves the axial currents in sect.~\ref{section:quarkmodel} 
have the same structure as Eq.(\ref{gp}), which when 
evaluated between 
SU(6) spin-flavor wave functions gives
\be
\label{gpga8}
\frac{g^8_{A}(0)}{g^8_{Aq}}= (a+2\, b). 
\ee

The advantage of a quark model calculation as the one presented here,
is that it provides expressions for these parameters in terms 
of the hypercharge axial coupling $g_{Aq}^8$ and a gluon exchange current
contribution, for example, $a \approx g_{Aq}^8$. Furthermore, 
we find $b<0$ because the matrix element of the hypercharge gluon 
exchange current operator in color space is negative. 

To calculate $g_A^8(0)$ absolutely, we need a numerical value
for the axial quark coupling $g_{Aq}^8$. We derive the latter from
the generalized Goldberger-Treiman relations in Eq.(\ref{gaqfeldmann}).
We use the following relations between the strong $\eta$ and $\eta'$ 
couplings to nucleons and quarks 
\bea
\label{strongcouprel}
\frac{g_{\eta N}}{M_N} & = & \frac{g_{\eta q}}{m_q} \, (A_1 + 2 \, A_2)  
\nonumber \\
\frac{g_{\eta' N}}{M_N} & = & \frac{g_{\eta' q}}{m_q} \,  (A_1 + 2\, A_2)  
\eea
and first determine $g_{\eta q}$ and $g_{\eta' q}$ from
the empirical $\eta N$ and $\eta' N$ coupling constants 
$g_{\eta N}$ and $g_{\eta' N}$, 
and then obtain $g^8_{Aq}$ and $g^0_{Aq}$ from Eq.(\ref{gaqfeldmann}).
Here, $A_1$ and $A_2$ denote one- and two-quark contributions to the
strong coupling constants. These are exactly the same as in 
Eq.(\ref{resultga8}). The reason for this exact correspondence is that 
the one- and two-quark operators needed to calculate the strong 
meson-baryon couplings are identical 
to the ones in the axial coupling calculation~\cite{Mor89,BH00}.

There are different experimental determinations for 
$g_{\eta N}$ ranging from $g_{\eta N}=2.1$~\cite{Zhu00,Kir98}
to $g_{\eta N}=2.75-4.6$~\cite{Pen87}. Flavor SU(3) symmetry predicts
$g_{\eta NN}=(3-4 \frac{D}{D+F})\, g_{\pi NN} \approx 3.6$, where
the $F$ and $D$ couplings determined from weak 
hyperon decays have been used~\cite{comment1}. 
For $g_{\eta' N}$ we are aware of two experimental determinations
from $\eta'$ photoproduction off nucleons giving 
$g_{\eta' N}=1.66$~\cite{Zha01} and $g_{\eta' N}=1.4\pm 0.1$~\cite{Dug06}. 
Taking the range of the experimental values for 
$g_{\eta N}$ and $g_{\eta' N}$ 
and the central values of the mixing parameters and decay 
constants from Table 1 in Ref.~\cite{Fel98} 
\bea 
\label{mixingparameters}
f_8 & = & (1.26 \pm 0.04)\, f_{\pi},  \qquad   
\Theta_8  =  -21.2^{\, \mathrm{o}} \pm 1.6^{\, \mathrm{o}} \nonumber \\
f_0 & = & (1.17 \pm 0.03) \, f_{\pi},  \qquad     
\Theta_0  =  -9.2^{\, \mathrm{o}} \pm 1.7^{\, \mathrm{o}}  
\eea
as input we obtain from Eq.(\ref{gaqfeldmann}) the following 
values for the axial quark couplings 
\bea
\label{qaxc}
g^8_{Aq} & = & 0.96 \pm 0.44  \nonumber \\
g^0_{Aq} & = & 0.46 \pm 0.09,
\eea
where the errors come from the experimental error bars on 
$g_{\eta N}$ and $g_{\eta' N}$. In addition, there are uncertainties
connected with the validity of the generalized Goldberger-Treiman 
relations and possible three-body axial currents which we have neglected.

\subsection{The flavor singlet axial coupling ${\bf g_A^0(0)}$}
\label{sect:sectionga0}

As discussed in sect. 2 the physical interpretation
of the flavor singlet axial coupling constant $g_A^0(0)$  
is the nucleon spin fraction 
carried by all quarks, i.e., valence plus sea quarks. 
To calculate $g_A^0(0)$ in the present model, we 
evaluate the matrix element of the sum of  
one- and two-quark flavor singlet axial current 
operators ${\bf A}^0({\bf q}\to 0)$ in sect.~\ref{section:quarkmodel} 
between proton wave functions~\cite{comment7}. We then find
\be
\label{resultga0}
\frac{g^0_{A}(0)}{g^0_{Aq}} = \frac{g^8_{A}(0)}{g^8_{Aq}}, 
\ee
i.e., the same result as in Eq.(\ref{resultga8}).

However, due to the existence
of the axial anomaly in the flavor singlet channel, 
U(3) symmetry is broken.
Consequently,  $g_{Aq}^0\ne g_{Aq}^8$ as 
reflected by the generalized Goldberger-Treiman relations 
of Eq.(\ref{gaqfeldmann}).
The ratio of the constituent quark axial couplings
$\zeta=g^0_{Aq}/g^8_{Aq}=0.48(24)$
obtained from Eq.(\ref{qaxc}) is positive. 
Thus, it is at variance with the negative $\zeta$
in Refs.~\cite{cheng}.
Also in view of Eq.(\ref{qcdcouplings}) a negative value
for either $g_{Aq}^8$ or $g_{Aq}^0$ appears to be 
rather unlikely.
Our result for $\zeta$ is in good agreement with the 
calculation in Ref.~\cite{Yab93} which gives $\zeta=0.53$. 
It would be interesting to calculate the hypercharge and singlet axial
quark couplings in the model
of Ref.~\cite{Peris}. 

Using Eq.(\ref{qaxc}) we get the results for the 
one-body and two-body axial current contributions to
$g_A^0(0)$ as shown in Table~\ref{table:flavorga80} (fourth row).
Due to the gluon exchange current and the renormalized flavor singlet 
axial quark coupling, the sum of quark spins in the nucleon 
is only about 1/3 of the value obtained in the additive quark model.
Thus it is possible to describe the singlet axial nucleon coupling 
in the framework of the chiral constituent quark potential model
provided consistent axial exchange currents are taken into account. 

Because of angular momentum conservation, this reduction of the quark spin 
is compensated by orbital angular momentum carried by the same
nonvalence quark degrees of freedom. This can be seen from the following
qualitative argument.
When a $u$-quark with positive spin projection emits a spin 1 gluon, 
its spin flips thereby reducing the fraction $\Delta u$ of the nucleon 
spin carried by the $u$-quarks.  The gluon moving in z-direction eventually 
annihilates into a quark-antiquark pair as in Fig.~\ref{fig:Fig1}(b). 
These sea quarks have momentum components transverse to the z-direction and
therefore contribute to the orbital angular momentum of the system.
In this way the gluon emission and annihilation process described by the
two-body exchange current of Fig.~\ref{fig:Fig1}(b)
leads to a redistribution of angular momentum from 
quark spin to orbital angular momentum carried by quark-antiquark 
pairs~\cite{note5}. 

This interpretation is supported by the results obtained from
pion electroproduction in the $\Delta$ resonance region, in particular
from the $N \to \Delta$ transition quadrupole moment.
The latter is a measure of the deviation of the nucleon charge density 
from spherical symmetry and therefore of the amount of 
orbital angular momentum in the nucleon. It was found that 
the $N \to \Delta$ quadrupole transition involves mainly
quark-antiquark degrees of freedom~\cite{BHF97,BH}, i.e., the 
same exchange current diagram of Fig.~\ref{fig:Fig1}(b) that leads 
here to the reduction of the quark spin contribution to the nucleon spin.
Thus, the electromagnetic $N\to \Delta$ quadrupole transition 
provides an independent indication that quark-antiquark degrees of 
freedom carry most of the orbital angular momentum in the nucleon.

\section{Summary}
\label{section:summary}

We have investigated the axial form factors related to the spin structure 
of the nucleon in the framework of the constituent quark model. 
In addition to the usual one-body axial current, we have included 
two-body axial currents as required by the PCAC condition. 
These two-body exchange currents provide an effective description 
of the nonvalence quark degrees of freedom 
in the nucleon. We have shown that they reduce the quark contribution 
to nucleon spin.

In particular, for the flavor octet hypercharge axial coupling $g_A^8(0)$, 
we obtain $g_A^8(0)=0.58\pm 0.26$ where the uncertainty 
comes mainly from the hypercharge axial quark coupling 
$g^8_{Aq}=0.96\pm 0.44$. The latter is determined from the experimental 
$\eta N$ and $\eta' N$ couplings via generalized Goldberger-Treiman relations
and reflects the experimental error bars of these coupling constants.
We emphasize that the reduction of the additive quark model 
$g_A^8(0)=1$ is predominantly due to the negative contribution of the 
quark-antiquark degrees of freedom effectively described by the axial 
gluon exchange current. 

Also in the case of the singlet axial coupling $g_A^0(0)$, 
measuring the quark and antiquark spin contribution to the 
nucleon spin, we find $g_A^0(0)=0.28 \pm 0.05$, i.e.,   
a drastic reduction compared to the standard additive quark model result
$g_A^0(0)=1$. This is partly due to
the negative contribution of the axial gluon exchange current, and
partly due to the renormalization of the 
singlet axial constituent quark coupling $g^0_{Aq}=0.46\pm 0.09$
as determined from the generalized Goldberger-Treiman relations.  
Both effects lead to a result in good agreement
with the experimental value $\Sigma =0.33(10)$. 

Thus, the failure of the
usual constituent quark model to describe the spin structure 
of the nucleon (``spin crisis'') is mainly a consequence of the incorrect
assumption that the axial nucleon current is the sum of three one-body quark
currents, underlying most applications of this model. We have pointed out
that this assumption is in conflict with the PCAC condition,
which demands that (i) the axial current operator contains two-body
axial exchange currents, and that (ii) the constituent quark axial
couplings are renormalized from their QCD quark values. 
In summary, including axial exchange currents
and assuming the validity of generalized Goldberger-Treiman relations 
allows a constituent quark model description of the hypercharge and singlet 
axial couplings of the nucleon that is consistent with the data.

This does not mean that all problems 
concerning nucleon spin structure are solved. 
There are important conceptual problems, in particular, 
the validity of the generalized  Goldberger-Treiman relations 
in the presence of the axial anomaly~\cite{Nap02,Che99,Nar99}, 
and the relativistic definition of spin itself~\cite{Sin98} 
which we have not addressed. Finally, it would be interesting 
to calculate the orbital angular momentum associated with the axial 
gluon exchange current investigated here.

\vspace{0.3 cm}
\centerline{{\bf ACKNOWLEDGMENTS}}
\noindent
Work supported in part 
by Spanish DGICYT under contract no. BFM2002-03218 and FPA2004-05616, 
and Junta de Castilla y Leon under contract no. SA104/04.


\begin{thebibliography}{99}

\bibitem{seh74} L. M. Sehgal, Phys. Rev. D {\bf 10}, 1663 (1974).
\bibitem{Ji} Xiangdong Ji, Phys. Rev. Lett. {\bf 78}, 610 (1997).
\bibitem{clo92} F. Close, Few-Body Systems, Suppl. {\bf 6}, 368 (1992).   
\bibitem{bro88} S. J. Brodsky, J. Ellis and M. Karliner, 
Phys. Lett. B{206}, 309, (1988); J. Ellis and M. Karliner, 
Phys. Lett. B{213}, 73, (1988). 
\bibitem{abe98} K. Abe et al. (E143 Collaboration), Phys. Rev. D {\bf 58},
112003 (1998).
\bibitem{emc} J. Ashman et al. (European Muon Collaboration),
Phys. Lett. B {\bf 206}, 364 (1988); Nucl. Phys. {\bf B328}, 1 (1989).
\bibitem{note0}
We use the notation 
$q_{\mu} = (\nu, -{\bf q})$, 
where $\nu$ is the energy transfer and ${\bf q}$ the three-momentum transfer 
to the proton.
\bibitem{carlitz} R. D. Carlitz, Int. J. Mod. Phys. E{\bf 1}, 505 (1992).
\bibitem{note1} In the following, we use the name ``QCD quark'' instead of 
``current quark'' to denote
the nearly massless quark fields appearing in the QCD Lagrangian.
\bibitem{note2} 
A description in which 
the axial current contains only one-body operators is often referred to as 
``impulse approximation''.
\bibitem{comment0}
Due to the axial gluon anomaly of QCD, gluon spin contributions 
$\Delta G(Q^2)$ are admixed
to the quark spin contributions in leading order perturbation
theory. As a result, the deep inelastic scattering experiments
actually measure $\Delta q(Q^2) = 
\tilde{\Delta q} -\alpha_S(Q^2) \Delta G(Q^2)$,
where $\alpha_S$ is the running QCD coupling constant.
Thus, the $Q^2$ dependence cancels in
the quark spin differences contained in $g_A(0)$ and $g_A^8(0)$ but 
remains in the quark spin sum $g_A^0(0)_{Q^2}$.
This $Q^2$ dependence is very soft in the perturbative regime, 
but its evolution down to the confinement scale is not known.
\bibitem{comment1} 
From neutron $\beta$-decay one can extract 
$g_A(0)=1.2670\pm 0.0035$~\cite{pdg}.
Similarly, from the $\beta$-decay of $\Xi^-$ hyperon,
and the assumption of SU(3) flavor symmetry~\cite{lipkin,flores} 
one obtains $g_A^8(0)=0.588\pm 0.033$
(see Ref.~\cite{abe98} and references therein).
Instead of the axial couplings $g_A(0)$ and $g_A^8(0)$, 
which govern the $\beta$-decay of octet baryons in the SU(3) limit, 
the symmetric and antisymmetric flavor octet 
coupling constants $D$ and $F$ are often used.
The relation between both notations is $g_A(0)= F+D$, and $g_A^8(0)= 3F-D$.
\bibitem{pdg} D.E. Groom et al. (Particle Data Group), Eur. Phys. J. C
{\bf 15}, 1 (2000). 
\bibitem{lipkin} H. J. Lipkin, Phys. Lett. B {\bf 214}, 429 (1988).
\bibitem{flores} R. Flores-Mendieta, E. Jenkins and A.V. Manohar, 
Phys. Rev. D {\bf 58}, 094028 (1996).
\bibitem{david} D. Barquilla-Cano, A.J. Buchmann 
and E. Hern\'andez, Nucl. Phys. {\bf A714}, 611 (2003).
\bibitem{BHY94} A. Buchmann, E. Hern\'andez, and K. Yazaki, 
Nucl. Phys. {\bf A569}, 661 (1994).
\bibitem{Eri88} For a derivation of the Goldberger Treiman relation
from the PCAC condition see: 
T. Ericson and W. Weise, Pions and Nuclei,
Clarendon Press, Oxford, 1988.
\bibitem{comment3}  
The pion-quark coupling constant $g_{\pi q}$ 
is fixed by the empirical pion-nucleon constant $g_{\pi N}$ via  
$g_{\pi q}=\frac{3}{5}\, \frac{m_q}{M_N} \, g_{\pi N} 
$. With 
$g_{\pi N}= 13.1$, $m_q=313$ MeV and $M_N=939$ MeV one obtains 
$g_{\pi q}=2.62$, and then from Eq.(\ref{gaq}), with $f_{\pi}=93$ MeV, 
$g_{Aq}=0.77$.
\bibitem{manohar} A. Manohar and H. Georgi, 
Nucl. Phys. {\bf B234}, 189 (1984). 
\bibitem{comment11}
The fields $\Phi_1$, $\Phi_2$, $\Phi_3$ correspond to the isovector 
$\pi$ fields, and $\Phi_8$ ($\Phi_0$)  
to the isoscalar $\eta_8$ ($\eta_0$) fields from which
the physical $\eta$ and $\eta'$ fields are constructed 
by mixing~\cite{nuclth}.
\bibitem{nuclth} D. Barquilla-Cano, A.J. Buchmann 
and E. Hern\'andez, Nucl. Phys. {\bf A 721}, 429c, (2003); nucl-th/0303020.
\bibitem{Yab93} H. Yabu, M. Takizawa, W. Weise, Z. Phys. A {\bf 345}, 193
(1993).
\bibitem{wei91} S. Weinberg,  Phys. Rev. Lett. {\bf 67}, 3473 (1991). 
\bibitem{Peris} S. Peris and E. de Rafael, Phys. Lett. B {\bf 309}, 
389 (1993); S. Peris, Phys. Lett. B {\bf 268}, 415 (1991); 
Phys. Rev. D {\bf 46}, 1202 (1992).
\bibitem{andreas} W. Broniowski, M. Lutz and A. Steiner, Phys. Rev. Lett.
{\bf 71}, 1787 (1993); 
U. Vogl, M. Lutz, S.Klimt and  W. Weise, Nucl. Phys. {\bf A516},
469 (1990). 
\bibitem{Leut98}H. Leutwyler, Nucl. Phys. B (Proc. Suppl.) {\bf 64},
 223 (1998); R. Kaiser, H.Leutwyler, hep-ph/9806336. 
\bibitem{Fel98} Th. Feldmann, Int. J. Mod. Phys. A{\bf 15}, 159 (2000);
Th. Feldmann, P. Kroll and B. Stech, Phys. Rev. D {\bf 58}
114006 (1998). 
\bibitem{Nap02} M. Napsuciale, A. Wirzba, M. Kirchbach, 
Nucl. Phys. {\bf A703}, 306 (2002).
\bibitem{Che99} T. P. Cheng, N. I. Kochelev, and V. Vento,
Mod. Phys. Lett. A {\bf 14}, 205 (1999).
\bibitem{Nar99} S. Narison, G. M. Shore, and G. Veneziano,
Nucl. Phys. {\bf B 546}, 235 (1999); G. Veneziano,
Mod. Phys. Lett. A {\bf 4}, 205 (1989), 347.
\bibitem{Mor89} G. Morpurgo, Phys. Rev. D {\bf 9}, 3111 (1989).
\bibitem{BH00} A. J. Buchmann and E. M. Henley, 
Phys. Lett. {\bf B 484}, 255 (2000). 
\bibitem{Zhu00} S.-L. Zhu, Phys. Rev. C {\bf 61}, 065205 (2000). 
\bibitem{Kir98} M. Kirchbach and H. J. Weber, 
Comments Nucl. Part. Phys. {\bf 22}, 171 (1998).
\bibitem{Pen87} J. C. Peng, Proc. of the LAMPF Workshop on Photon and Neutral
Meson Physics at Intermediate Energies-LA-11177-C, Los Alamos, NM, Jan. 7-9, 
1987, edited by H. W. Baer et al. (Los Alamos National Laboratory 1987).
\bibitem{Zha01} Q. Zhao, Phys. Rev. C {\bf 63}, 035205 (2001).
\bibitem{Dug06} M. Dugger, J.P. Ball, P. Collins, E. Pasyuk, B.G. Ritchie
et al., Phys. Rev. Lett. {\bf 96}, 062001 (2006). 
\bibitem{comment7} 
Although the evolution of $g_A^0(0)_{{\mu}^2}$ into the confinement 
region is not known we compare our quark model result
with data taken at the renormalization scale $\mu^2=Q^2=3 (GeV/c)^2$. 
\bibitem{cheng} T. P. Cheng and L.-F. Li, Phys. Rev. Lett. {\bf 74},
2872 (1995); X. Song, Int. J. Mod. Phys. A {\bf 16}, 3673 (2001) \ 
hep-ph/0012295; H. Dahiya and M. Gupta, Int. J. Mod. Phys. A {\bf 19}, 
5027 (2004) \ hep-ph/0305327.
\bibitem{note5} 
A similar conclusion concerning the redistribution 
from quark spin to orbital angular momentum was reached
in Ref.~\cite{cheng}.
\bibitem{BHF97} 
A. J. Buchmann, E. Hern\'andez, and A. F\"a\ss ler, 
Phys. Rev. C {\bf 55}, 448 (1997).
\bibitem{BH} A. J. Buchmann and E. M. Henley, Phys. Rev. C {\bf 63}, 
015202 (2001).  
\bibitem{Sin98} D. Singleton, Phys. Lett. {\bf B427}, 155 (1998).
\end{thebibliography}
\end{document}